# Large Unidirectional Magnetoresistance in Metallic Heterostructures in the Spin Transfer Torque Regime

Ting-Yu Chang[1], Chih-Lin Cheng[1], Chao-Chung Huang[1], Cheng-Wei Peng[1], Yu-Hao Huang[1], Tian-Yue Chen[1], Yan-Ting Liu[1,†], and Chi-Feng Pai[1,2,*]

[1]*Department of Materials Science and Engineering, National Taiwan University, Taipei 10617, Taiwan*

[2]*Center of Atomic Initiative for New Materials, National Taiwan University, Taipei 10617, Taiwan*

A large unidirectional magnetoresistance (UMR) ratio of UMR/$R_{xx}$ ~ 0.36% is found in W/CoFeB metallic bilayer heterostructures at room temperature. Three different regimes in terms of the current dependence of UMR ratio are identified: A spin-dependent-scattering mechanism regime at small current densities $J$ ~ $10^9$ A/m$^2$ (UMR ratio $\propto J$), a spin-magnon-interaction mechanism regime at intermediate $J$ ~ $10^{10}$ A/m$^2$ (UMR ratio $\propto J^3$), and a spin-transfer torque (STT) regime at $J$ ~ $10^{11}$ A/m$^2$ (UMR ratio independent of $J$). We verify the direct correlation between this large UMR and the transfer of spin angular momentum from the W layer to the CoFeB layer by both field-dependent and current-dependent UMR characterizations. Numerical simulations further confirm that the large STT-UMR stems from the tilting of the magnetization affected by the spin Hall effect-induced spin-transfer torques. An alternative approach to estimate damping-like spin torque efficiencies from magnetic heterostructures is also proposed.

† Email: f07527076@ntu.edu.tw
* Email: cfpai@ntu.edu.tw

# I. INTRODUCTION

Magnetoresistance (MR) is one of the key physical properties to understand spin-dependent transport mechanisms in various condensed matter systems. The discovery of the giant magnetoresistance (GMR) [1-3] in ferromagnetic-normal metal (FM/NM) multilayer structures as well as giant tunneling magnetoresistance (TMR) in MgO-based magnetic tunnel junctions (MTJs) [4-6] further revolutionized the hard-disk drive and the magnetic memory technologies, respectively. More recently, the spin Hall effect (SHE) [7-9] in magnetic heterostructures with strong spin-orbit interactions is also found to give rise to novel MRs, namely the spin Hall MR (SMR) [10-12] and the unidirectional SMR (USMR) [13-16]. These newly-discovered MR effects can not only be utilized to quantify charge-to-spin conversion efficiencies in various types of NM/FM magnetic heterostructures but also serve as alternate readouts for emergent magnetoresistive memory devices. From the origin point of view, the theoretical framework for SMR is more well-established, in which the MR stems from the absorption or the reflection of the SHE-induced spin current at the NM/FM interface and thereby affecting the longitudinal resistance of the heterostructure via the inverse spin Hall effect (ISHE) [10,17]. Typical longitudinal SMR ratio ranges from $\Delta\rho_{xx}/\rho_{xx}$ (or $\Delta R_{xx}/R_{xx}$) ~ 0.001% to 1.3% [10-12]. In contrast, a solid theoretical model for USMR or UMR is still lacking, although its origin is typically attributed to an interfacial spin-dependent scattering scenario (similar to the GMR case [1,2]), with typical UMR ratio ranges from ~ 0.0019% to 1.1% [13,14,16,18]. It is later found that the UMR in NM/FM heterostructures consists of two major components, a spin-dependent scattering contribution (could be of bulk and/or interfacial origins) [13,14,16] and an electron-magnon scattering contribution [19-21], as respectively illustrated in Fig. 1 (a) and (b).

In this work, we disentangle the underlying mechanisms of the UMR in NM/FM (W/CoFeB) magnetic heterostructures by systematic current-dependent and field-dependent MR measurements at room temperature. A broad range of current densities are applied, ranging from ~ $10^9$ A/m$^2$ (AC signal) to ~ $10^{11}$ A/m$^2$ (DC signal). A third regime of the UMR is found to be directly related to the damping-like spin torque (DL-ST) efficiency $\xi_{DL}$ of the magnetic heterostructures, which is similar to the SMR case [11,12,22]. The UMR of W/CoFeB heterostructures will first increase linearly and then nonlinearly to the applied current density $J$ as $J$ reaches ~ $5\times10^{10}$ A/m$^2$, which corresponds to a later defined inflection current density $J_{inflection}$. The UMR of W/CoFeB will further saturate at a UMR ratio of ~ 0.36% as $J$ reaches ~ $10^{11}$ A/m$^2$. By performing macrospin and micromagnetic simulations, the current-induced STs from the SHE of the NM layer is confirmed to cause the tilting of the FM magnetization thereby creating a change in MR. Consequently, we define this additional UMR as the spin-transfer torque UMR (STT-UMR), as illustrated in Fig. 1 (c). Meanwhile, $\xi_{DL}$ of the tested heterostructures can be roughly estimated by using both the experimental UMR data and the simulation results.

## II. MATERIALS SYSTEM

A series of W($t_W$)/Co$_{40}$Fe$_{40}$B$_{20}$(2)/MgO(1)/Ta(2) ($t_W$ is the thickness of the W layer and numbers in parentheses are in nm) multilayer heterostructures are deposited onto Si/SiO$_2$ substrates by high-vacuum magnetron sputtering (base pressure ~ 10$^{-8}$ Torr) with working Ar pressures of 3 mTorr (10 mTorr) for DC (RF) sputtering. The thickness of the W layer ranges from $t_W$ = 2 nm to 7 nm. The top Ta layer serves as capping layer to prevent other layers from oxidation [23]. Here, all samples show in-plane magnetic anisotropy. Saturation magnetization of the CoFeB layer is $M_s \approx$ 700 emu/cm$^3$, as characterized by vibrating sample magnetometer (VSM). Besides, the thickness of the magnetic dead layer is negligible in these as-deposited films [24]. To perform MR measurements, thin films are patterned into micron-sized Hall bar devices with channel width of 5 μm through photolithography and lift-off processes.

## III. ORIGINS OF THE UMR

As shown in Fig. 2(a), to measure longitudinal resistance ($R_{xx}$) on Hall bar devices, we sweep the in-plane magnetic fields along *y*-direction ($H_y$) while applying DC (opposite pulsed currents with 0.5 s duration) or AC currents (frequency $\omega/2\pi$ = 83 Hz) along *x*-direction [13,24]. We use a source meter (Keithley 2400) and a lock-in amplifier (Signal Recovery 7265) to supply DC currents and AC currents, respectively. As preliminary tests, we sweep the in-plane magnetic fields within a smaller field range ($H_{y,\text{max}} = \pm 600$ Oe). Representative normalized UMR loops of W(4)/Co$_{40}$Fe$_{40}$B$_{20}$(2) with AC current (amplitude) $I_{\text{sense}} = 0.079$ mA and DC currents $I_{\text{sense}} = \pm 1$ mA are shown in Fig. 2(b) and Fig. 2(c), respectively. For the DC measurement, we record the difference of longitudinal resistance $\Delta R = R_{xx}^{I+} - R_{xx}^{I-}$ for applying opposite directions of DC. Note that the protocol for the DC measurement is equivalent to the Fourier-transformed second harmonic ($2\omega$) signals in the AC measurement. We define the UMR as $|[(R_{H+}^{I+} - R_{H+}^{I-}) - (R_{H-}^{I+} - R_{H-}^{I-})]/2|$ and $|R_{2\omega}^{H+} - R_{2\omega}^{H-}|$ for DC and AC measurements, respectively [13,24,25]. From these experimental results, we observe that the UMR measured by DC currents is much larger than that by an AC current, suggesting an influence of the applied current magnitude in the UMR.

Subsequently, to examine the current-dependence of the UMR, different amplitudes of DC and AC currents are applied to the W(4)/CoFeB(2) sample. The DC and AC current amplitudes are collectively called sense currents ($I_{\text{sense}}$) in this section. As shown in Fig. 2(d), the UMR ratio (UMR/$R_{xx}$) with applying AC currents is linear to $I_{\text{sense}}$, which is consistent with previous reports in such low current density regime [13,16,26]. UMR/$R_{xx}$ from both DC and AC measurements are shown together in Fig. 2(e), which can be roughly divided into three regimes: In the first regime with a lower $I_{\text{sense}}$, UMR/$R_{xx}$ increases linearly. In the second and the third regimes, UMR/$R_{xx}$ rises rapidly (nonlinearly) and reaches a saturated value (~ 0.36%) with increasing $I_{\text{sense}}$.

As previous works reported, conventional UMR can be separated into two major components, namely, the spin-dependent UMR (SD-UMR) and the spin-flip UMR (SF-UMR) [16,26]. These two types of UMR come from two different competing mechanisms: The SD-UMR originated from spin-dependent-scattering at a large field and low current regime while the SF-UMR is attributed to electron-magnon scattering at a low field and large current regime. Since the spin-dependent-scattering mechanism is only related to the amount of spin accumulation at the NM/FM interface and the orientation of FM magnetization **M** with respect to *y*-direction [Fig. 1(a)] [13-15], SD-UMR is proportional to the magnitude of the applied current and independent of the external field, which can be described as $aI$ ($a$ is a field-independent coefficient). For the SF-UMR resulting from electron-magnon scattering [Fig. 1(b)], the variation of thermal-induced and spin current-induced magnons causes the nonlinear current-dependent behavior of the UMR, which shows a $bI+cI^3$ trend with the magnetic field-dependent coefficients $b$ and $c$ [16,26-30].

As shown in Fig. 2(e), the UMR/$R_{xx}$ of the W(4)/CoFeB(2) sample ranging from 0.06 to 1.2 mA (corresponds to current densities $J = 2.1\times10^9$ to $4.0\times10^{10}$ A/m$^2$) is in agreement with the trend of $(a+b)I+cI^3$, which suggests the behavior of the UMR can be explained by those mechanisms mentioned above. Followed by the fitting line, we can further separate this part of UMR/$R_{xx}$ into two regimes: At small $I_{\text{sense}}$ measured by applying AC, UMR/$R_{xx}$ is linearly proportional to $I_{\text{sense}}$ [Fig. 2(d)], which comes from the SD-UMR. After that, it increases nonlinearly with the increase of $I_{\text{sense}}$ due to the SF-UMR. However, as $I_{\text{sense}}$ becomes larger than 1.2 mA ($J = 4.0\times10^{10}$ A/m$^2$), the trend of UMR vs. $I_{sense}$ deviates from both the SD-UMR and the SF-UMR trends, suggesting an additional mechanism emerging at the high current regime. Besides, $I_{sense}$ vs. UMR/$R_{xx}$ is found to have an inflection point at $I_{sense}$ = 1.6 mA, signaling that the dominating mechanism has changed near this point, and we define this current as $I_{inflection}$ ($J_{inflection} \sim 5.0 \times 10^{10}$ A/m$^2$).

To further gain insight into this additional UMR contribution, we measure UMR$_{sat}$/$R_{xx}$ in a series of W($t_W$)/CoFeB(2) samples with $H_{y,max}$ = ±600 Oe. As shown in Fig. 2(f), we observe that UMR$_{sat}$/$R_{xx}$ reaches a maximum (~ 0.36%) at $t_W$ = 4 nm and then proceeds to decrease as further increasing $t_W$, which is similar to the trend of DL-ST efficiency $\xi_{DL}$ vs. W thickness due to the phase transition from amorphous W to α-W with increasing $t_W$ [22,31-33]. Note that $\xi_{DL}$ is related to the internal spin Hall ratio (spin Hall angle) $\theta_{SH}$ of the NM layer through $\xi_{DL} = \theta_{SH} T_{int}$, which describes the apparent efficiency of the charge-to-spin conversion. $T_{int}$ is the spin transparency of the NM/FM interface [34]. Consequently, we believe that this additional UMR is related to the SHE-induced ST transfer from the heavy metal layer W layer into the CoFeB layer and we tentatively define this extra UMR as the spin-transfer torque UMR (STT-UMR). It is also worth noting that the anomalous Nernst effect and the spin Seebeck effect can also give rise to a similar longitudinal resistance. Nevertheless, such thermal-induced signal is typically ~ 1 mΩ in metallic systems [13,15,35], which is three orders of magnitude smaller than the UMR observed in the STT-dominated regime in our samples, therefore resulting in a comparably negligible thermal contribution.

In Table. 1, we compare the maximum UMR ratio of our W/CoFeB heterostructures with different materials systems previously reported. Since the maximum UMR ratio in this report is independent of the applied current at the large current regime, we take other groups' UMR ratios under the largest current applied for a fair comparison. Note that although the maximum UMR ratio for the W/CoFeB heterostructure is only slightly smaller than those for GaMnAs/BiSb and $Cr_x(Bi_{1-y}Sb_y)_{2-x}Te_3/(Bi_{1-y}Sb_y)_2Te_3$(CBST/BST) heterostructures involving topological insulators (under cryogenic condition), it can be detected at room temperature and is much larger than those observed in other NM/FM bilayer systems, *e.g.*, W/Co or Pt/Co [13,14,16,18,26], potentially due to a more pronounced ST contribution. Other DC current-induced MR effect has also been observed in $Ga_{0.91}Mn_{0.09}As/ Ga_{0.97}Mn_{0.03}As$ bilayer system, which is coined as the linear spin Hall magnetoresistance (LSMR) [25]. However, since the LSMR mainly comes from the thermal effect in such semiconductor system, it is not included here for comparison.

## IV. CURRENT-DEPENDENT AND FIELD-DEPENDENT UMR

We further perform field-dependent UMR measurements with a wider field range (up to 5 k Oe) while applying various $I_{sense}$ to investigate the possible origins of such STT-UMR. As shown in Fig. 3(a), for a W(4)/CoFeB(2) sample, the trend of field-dependent UMR/$R_{xx}$ with $I_{sense}$ = 0.8 mA ($J$ = 2.7×10$^{10}$ A/m$^2$) follows the fitting line of $H_y^{-p}$ with the exponent $p$ = 1.37, which indicates that the UMR/$R_{xx}$ at small current densities is indeed governed by the SF-UMR [16].

With the increase of $I_{sense}$, the effect of ST begins to emerge and the field-dependent UMR/$R_{xx}$ no longer follows $H_y^{-p}$. Since the SF-UMR are suppressed by the ST, the relative SF-UMR contribution will decrease, causing the deviation of the field-dependent UMR/$R_{xx}$ from the trend of $H_y^{-p}$. For $I_{sense}$ = 1.6 mA ($J$ = 5.4×10$^{10}$A/m$^2$), the field-dependent UMR/$R_{xx}$ shows a linear trend at the low field regime, suggesting that the competition between the SF- and the STT-UMR has begun. As $I_{sense}$ gets even larger, STT-UMR becomes the dominating mechanism of the measured UMR, and the UMR ratio remains fairly constant (~ 0.36%) at the low $H_y$ regime. However, regardless of $I_{sense}$, the STT-UMR will eventually be canceled out at a sufficiently large $H_y$, indicating that the current-induced ST or its effective field is suppressed by $H_y$.

As an alternative way to demonstrate this effect, we measure UMR/$R_{xx}$ vs. $I_{sense}$ with various $H_y$. As shown in Fig. 3(b), there are two major features: First, the saturated UMR$_{sat}$/$R_{xx}$ decreases from 0.36% to 0.09% with the increase of $H_y$. Second, $I_{inflection}$ becomes larger as $H_y$ increases, which suggests that a larger ST is required to overcome the increase of $H_y$. Based on the field-dependent and the current-dependent UMR/$R_{xx}$, we find that both trends are in disagreement with the fitting curves at the low field ($H_y^{-p}$) and in the high current regime [$(a+b)I+cI^3$]. These results suggest that SF-UMR no longer governs UMR/$R_{xx}$ and STT-UMR dominates UMR/$R_{xx}$ in the high current regime.

In Fig. 3(c) we summarize the $H_y$ dependence of the inflection current density $J_{inflection}$, where a linear trend is found with a slope of $J_{inflection}$ / $\mu_0 H_y$ ~ 2.14×10$^{11}$A/(m$^2$ T). These features again indicate that the extra UMR is a result of the competition between ST and $H_y$, since $H_y$ aligns **M** towards *y*-direction whereas **M** can also be tilted towards *x* and *z*-directions by applying a countering ST. Therefore, we further suspect that this additional UMR originated from the SMR ($\propto 1-m_y^2$) and/or anisotropic magnetoresistance (AMR, $\propto m_x^2$) caused by the ST-induced tilting of **M** [13] in the heterostructure [Fig. 1(c)].

## V. NUMERICAL SIMULATIONS

To confirm that STT-UMR can be attributed to the extra SMR and/or AMR contributions due to the competition between ST and $H_y$, we perform macrospin simulations to investigate the influence of current-induced STs (both DL and field-like, FL) and applied fields on the magnetization $\mathbf{M} = (m_x, m_y, m_z)$. The macrospin simulations are based on LLG equation with additional ST terms:

$$\frac{d\mathbf{m}}{dt} = -\gamma\, \mathbf{m}\times\mathbf{H_{eff}} + \alpha\, \mathbf{m}\times\frac{d\mathbf{m}}{dt} + \gamma\, H_{DL}\mathbf{m}\times(\boldsymbol{\sigma}\times\mathbf{m}) + \gamma\, H_{FL}(\boldsymbol{\sigma}\times\mathbf{m}), \tag{1}$$

where $\mathbf{H_{eff}}$ is effective field composed of external field and anisotropy field, $\gamma$ is the gyromagnetic ratio, $\alpha$ is the Gilbert damping constant, $H_{DL}$ and $H_{FL}$ are the effective fields originated from the current-induced damping-like ST (DL-ST) and field-like ST (FL-ST) generated by the SHE in the heavy metal layer. We consider both effective fields are proportional to the applied current density, $J_{sense}$. Damping constant $\alpha$ and effective out-of-plane anisotropy field ($\mu_0 H_{K,out}$) are set to be 0.03 and 0.44 T, respectively, which are obtained experimentally by ferromagnetic resonance measurements using our W/CoFeB samples. The orientation of spin polarization ($\sigma$) is set along $y$-direction to compete with $H_y$. We set the duration of STs and $H_y$ as 120 ns and record the time-average $y$-component of magnetization ($m_{y,avg}$) for the last 40 ns to observe this competing effect. Since both effective fields are proportional to the applied current, $H_{FL}/H_{DL}$ is set at a constant 0.2 for generality and to take the FL-ST into account.

We first observe that the current-dependent variation of $m_x$ is limited, therefore the AMR contribution ($\propto m_x^2$) can be ruled out and reveals that the additional UMR mainly stems from the SMR term ($\propto 1-m_y^2$). As shown in Fig. 4(a), we sweep $\mu_0 H_{DL}$ from 0 mT to 70 mT (corresponds to increasing $J_{sense}$ and therefore the DL-ST) to observe the variation of $m_{y,avg}$ at different $\mu_0 H_y$ (with $\sigma$ opposing $H_y$). The DL-ST from $\sigma$ that is opposite to the external field will compete with $H_y$ and deviate the magnetization away from the $y$-direction. As increasing $\mu_0 H_{DL}$ from 0 mT to a specific value, the magnetization orientation can be tilted (a jump in $m_{y,avg}$), which corresponds to the DL-ST switching. This specific value becomes larger as $\mu_0 H_y$ increases and we define this threshold value as $\mu_0 H_{DL,th}$. As shown in Fig. 4(b), we plot $\mu_0 H_{DL,th}$ as a function of $\mu_0 H_y$ and the trend can be well fitted by a linear function with a slope $H_{DL,th}/H_y \sim 0.03$. Assuming that the numerically determined $H_{DL,th}$ corresponds to the experimentally determined $J_{inflection}$, we can further estimate the DL-ST efficiency of the W(4)/CoFeB(2) device by [36,37]

$$\xi_{\mathrm{DL}} \approx \frac{2e\mu_0 M_s t_{\mathrm{FM}}}{\hbar}\left(\frac{H_{\mathrm{DL,th}}}{J_{\mathrm{inflection}}}\right) = \frac{2eM_s t_{\mathrm{FM}}}{\hbar}\left(\frac{H_{\mathrm{DL,th}}/H_y}{J_{\mathrm{inflection}}/\mu_0 H_y}\right), \qquad (2)$$

from which $\xi_{\mathrm{DL}} \sim 0.59$ is determined. Another observable feature is that in the large $\mu_0 H_{\mathrm{DL}}$ regime, the saturated $m_{y,\mathrm{avg}}$ increases as increasing $H_y$, which will result in a lower SMR contribution ($\propto 1-m_{y,\mathrm{avg}}^2$). This trend is also consistent with the experimental observation of a lower saturated UMR under a larger $H_y$, as shown in Fig. 3(b).

To see how different factors could affect the simulation and thereby the efficiency estimation result, we perform additional macrospin simulations by varying parameters that can influence ST-driven magnetization switching, such as damping constant $\alpha$, anisotropy field $\mu_0 H_{k,\mathrm{out}}$ and FL-ST/DL-ST ratios. As shown in Fig. 4(c) and (d), different magnitudes of $\alpha$ can affect the slope $H_{\mathrm{DL,th}}/H_y$, which increases linearly to $\alpha$. This suggests that an accurate determination of $\alpha$ experimentally is important for using this protocol to estimate $\xi_{\mathrm{DL}}$. We also extract $H_{\mathrm{DL,th}}/H_y$ with various $\mu_0 H_{k,\mathrm{out}}$ and FL-ST/DL-ST ratios, which show almost the same slope (Fig. 4(e) and (f)), suggesting that the magnitude of $\mu_0 H_{k,\mathrm{out}}$ and the existence of FL-ST play minor roles in affecting the estimation. In short, the value of $\alpha$ will influence the result of the estimated $\xi_{\mathrm{DL}}$ in this method, while the variation of $\mu_0 H_{k,\mathrm{out}}$ and FL-ST will not.

We further perform micromagnetic simulations via Ubermag [38] (The calculation kernel is based on OOMMF [39]) and Mumax3 [40] to see if the geometry of the simulated device could affect the estimation. In both Ubermag and Mumax3 simulations, the device geometry is set as 100 nm×100 nm×2 nm, which is split into 400 cells with dimensions of 5 nm×5 nm×2 nm. Note that this is actually much smaller than the actual device for our experimental tests. To consider exchange and Dzyaloshinskii-Moriya interactions (DMI) into $\mathbf{H_{eff}}$, the exchange stiffness constant and the interfacial DMI is set as $1.6\times10^{-11}$ J/m and $2\times10^{-4}$ J/m$^2$, respectively. The strength of $\xi_{DL}$ is set as 0.5 for simplicity. Other parameters are set to be the same as the macrospin simulations. The simulated critical current density $J_{\text{inflection}}$ is defined as the threshold current density to tilt the magnetization orientation by ST, which has the same physical meaning as the above mentioned $\mu_0 H_{DL,th}$. The $\mu_0 H_y$ dependence of $J_{\text{inflection}}$ exhibits the same linear trend as found in experiments and macrospin simulations. The slope, $J_{\text{inflection}}/\mu_0 H_y \sim 2.20\times10^{11}$ A/(m$^2$ T), is fairly consistent with the experimental results shown in Fig. 3(c).

## VI. ESTIMATION OF THE DL-ST EFFICIENCIES

The protocol mentioned above allows us to systematically determine $\xi_{DL}$ from a series of W($t_W$)/CoFeB(2) devices. As shown in Fig. 5(a), $J_{\text{inflection}}/\mu_0 H_y$ decreases from $2.33\times10^{11}$ A/(m$^2$T) at $t_W$ = 2 nm to $2.14\times10^{11}$ A/(m$^2$T) at $t_W$ = 4 nm, and then increases to $2.54\times10^{11}$ A/(m$^2$T) at $t_W$ = 7 nm. By employing $J_{\text{inflection}}/\mu_0 H_y$ (from experiments) and $H_{DL,th}/H_y$ (from simulations) data, the thickness dependence of $\xi_{DL}$ is estimated by Eqn. (2) and summarized in Fig. 5(b). $\xi_{DL}$ reaches ~ 0.59 at $t_W$ = 4 nm and then proceeds to decrease as W thickness increases. This phase transition behavior of $\xi_{DL}$ in the W/CoFeB heterostructures is fairly consistent with previous reports [22,32,33]. However, since the size of our sample is in micron-meter regime and the proposed estimation protocol is mainly related to the switching process under $H_y$ and ST, some uncertainties in estimation of $\xi_{DL}$ may arise due to the oversimplified macrospin model. This might lead to an overestimation in $\xi_{DL}$, since the switching behavior in larger size samples typically involves multidomain nucleation therefore could deviate from the macrospin or single domain prediction [41]. We believe that a more accurate efficiency estimation can be achieved either by shrinking size of the tested device or by performing micromagnetic simulations using a larger device size.

## VII. CONCLUSION

To conclude, through systematic current-dependent and field-dependent UMR measurements, we discover an additional ST-induced UMR (STT-UMR) at the high current regime, which leads to a large UMR ratio of ~ 0.36% at room temperature for W/CoFeB magnetic heterostructures. This STT-UMR can be attributed to an extra contribution from SMR, which originates from the tilting of magnetization **M** caused by the competition between the DL-ST effective field $H_{DL}$ and the applied external field $H_y$. This is confirmed by both macrospin and micromagnetic simulations, from which the numerically determined $H_{DL,th}/H_y$ can be employed to estimate $\xi_{DL}$ together with the experimentally obtained $J_{inflection}/\mu_0 H_y$. Our studies thus confirm the correlation between the SHE-induced ST and the large room-temperature UMR, thereby providing an alternative approach to characterize DL-ST efficiency.

## Acknowledgements

The authors acknowledge support from the Ministry of Science and Technology of Taiwan (MOST) under grant No. MOST-110-2636-M-002-013. The work is also partly supported by Taiwan Semiconductor Manufacturing Company (TSMC).

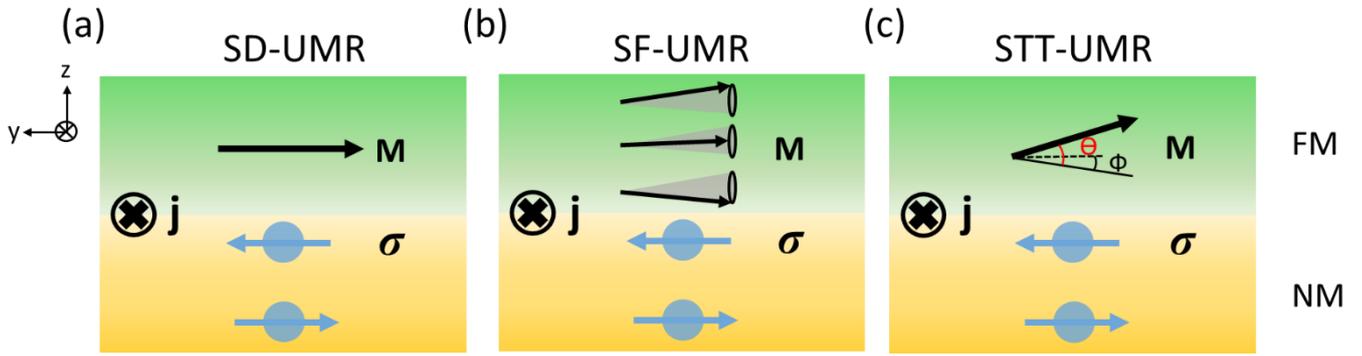

FIG. 1. Illustration of (a) spin-dependent UMR (SD-UMR), (b) spin-flip UMR (SF-UMR), and (c) spin-transfer torque UMR (STT-UMR).

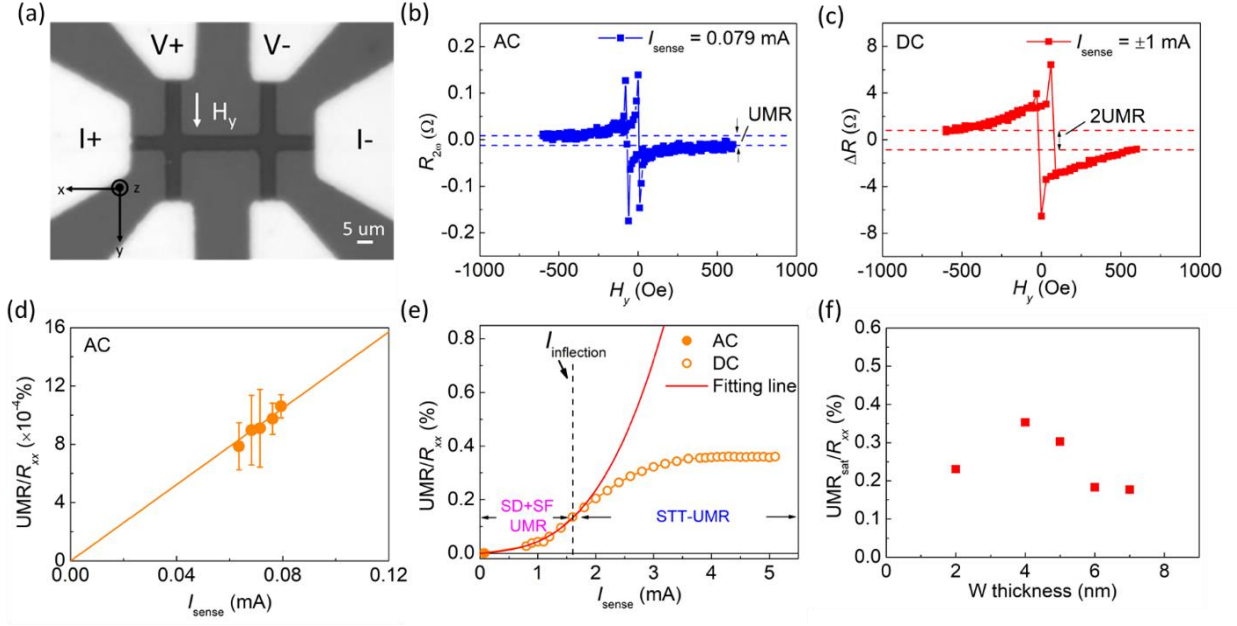

FIG. 2. (a) Optical microscopic (OM) image of a W/CoFeB Hall bar device for UMR measurement, which is performed with applying either AC or DC currents. Representative UMR loops as functions of in-plane field $H_y$ with (b) a AC current (amplitude) $I_{sense}= 0.079$ mA and (c) DC currents $I_{sense}=\pm 1$ mA for a W(4)/CoFeB(2) sample. (d) UMR/$R_{xx}$ versus $I_{sense}$ of a W(4)/CoFeB(2) device measured by AC currents with $H_{y,max}=\pm 600$ Oe. The solid line represents a linear fit to the experimental data. (e) Whole range $I_{sense}$ dependence of UMR/$R_{xx}$ with $H_{y,max}=\pm 600$ Oe for a W(4)/CoFeB(2) sample. The red solid line is a fit to $(a+b)I+cI^3$ for data up to the inflection current $I_{inflection}$. (f) UMR/$R_{xx}$ of W($t_W$)/CoFeB(2) samples as a function of W thickness ($t_W$) with $H_{y,max}=\pm 600$ Oe.

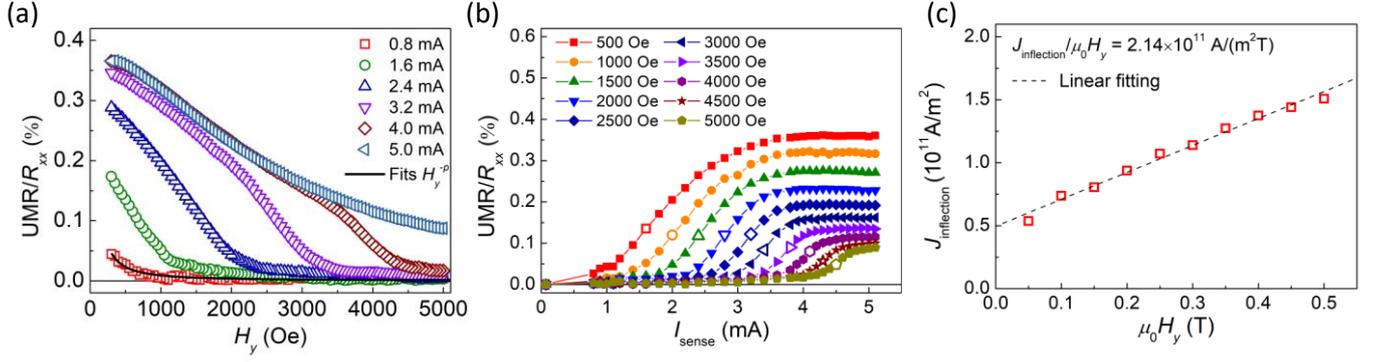

FIG. 3. (a) Field-dependent UMR/$R_{xx}$ of a W(4)/CoFeB(2) sample with various $I_{sense}$. The solid line represents a fit to the experimental data of $I_{sense}$ = 0.8 mA with the exponent $p$ = 1.37. (b) Current-dependent UMR/$R_{xx}$ with different $H_y$. The open dots in each set of data represent $I_{inflection}$ under different $H_y$. (c) $\mu_0 H_y$ dependence of $J_{inflection}$ extracted from Fig. 3(b). The dashed line is a linear fit to the data.

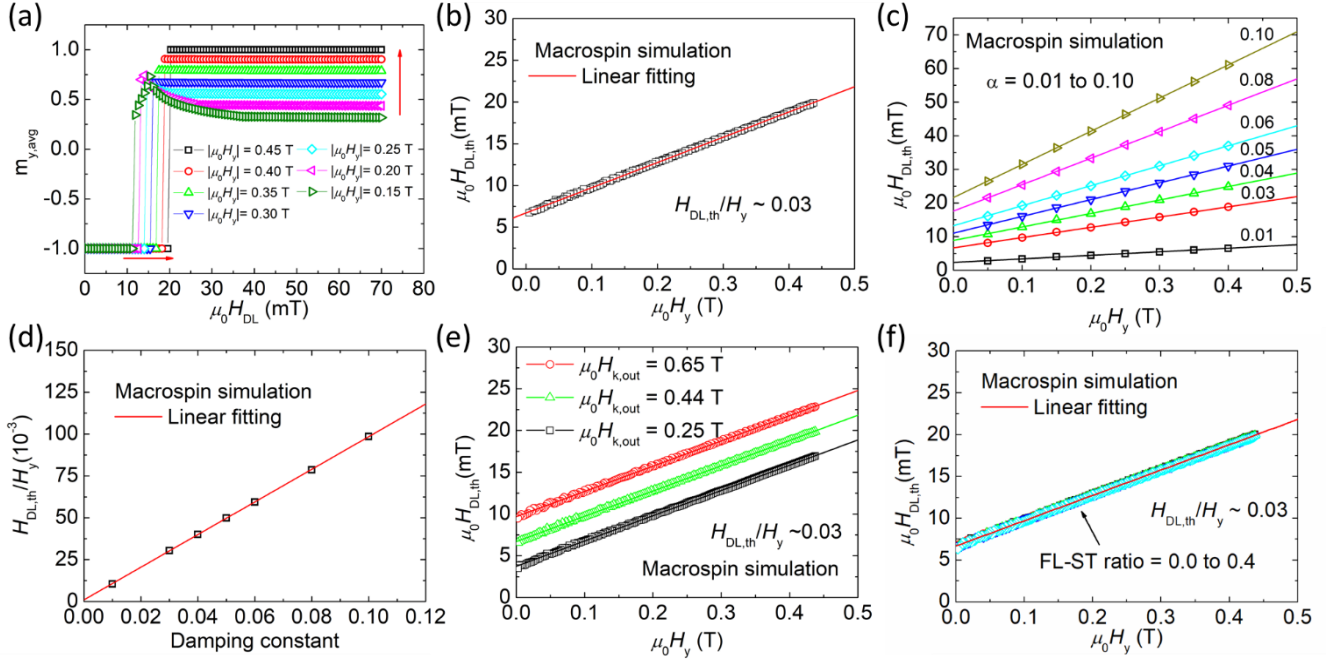

FIG. 4 Macrospin simulation: (a) $m_{y,avg}$ as a function of $\mu_0 H_{DL}$ ($\propto J_{sense}$) with different $\mu_0 H_y$. The red arrows represent the variation of $\mu_0 H_{DL,th}$ and $m_{y,avg}$ with increasing $H_y$. (b) $\mu_0 H_y$ dependence of $\mu_0 H_{DL,th}$ with FL-ST/DL-ST ratio = 0.2. (c) $\mu_0 H_y$ dependence of $\mu_0 H_{DL,th}$ with damping constant $\alpha$ = 0.01 to 0.10. (d) The summary of slope $H_{DL,th}/H_y$ vs. damping constant, which are extracted from (a). (e) $\mu_0 H_y$ dependence of $\mu_0 H_{DL,th}$ with $\mu_0 H_{k,out}$ = 0.25 T, 0.44 T and 0.65 T. (f) $\mu_0 H_y$ dependence of $\mu_0 H_{DL,th}$ with FL-ST/DL-ST ratio = 0.0 to 0.4. The results for various FL-ST/DL-ST ratios are almost overlapped with each other.

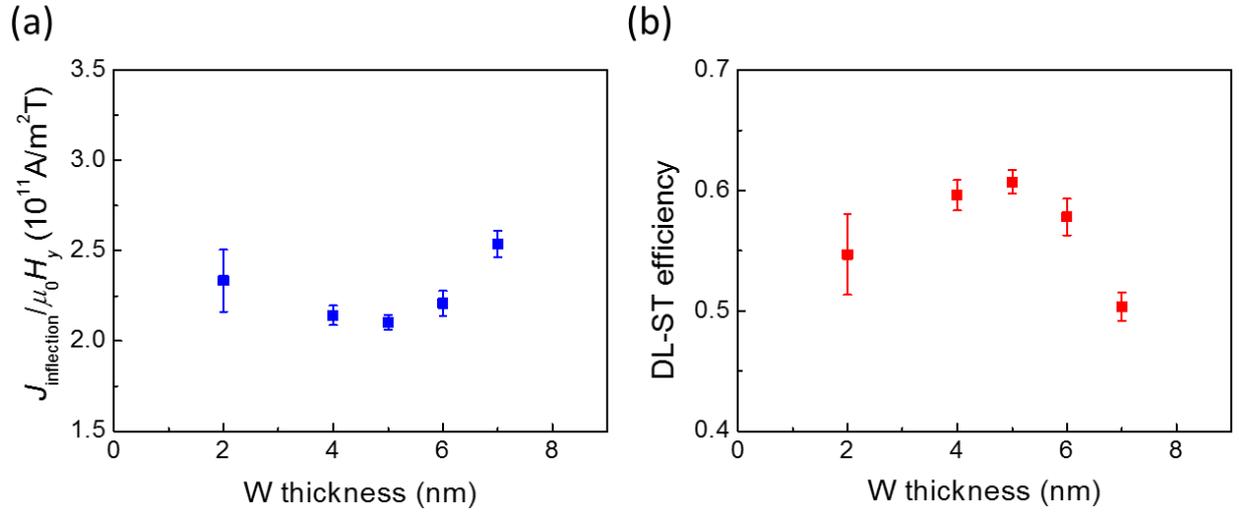

FIG. 5. (a) Experimentally obtained $J_{\text{inflection}} / \mu_0 H_y$ and (b) the estimated DL-ST efficiencies $\xi_{\text{DL}}$ of W($t_W$)/CoFeB(2) samples as functions of W thickness.

TABLE I. Comparison of the UMR magnitude from different materials systems.

| Materials system | $J$ (A/m$^2$) | UMR ratio (%) (maximum) | Field range (T) | Reference |
| --- | --- | --- | --- | --- |
| W/CoFeB (RT) | $1.67 \times 10^{11}$ | 0.36 | 0.06 | This work |
| W/CoFeB (RT) | $1.67 \times 10^{11}$ | 0.1 | 0.5 | This work |
| W/Co (RT) | $10^{11}$ | 0.0019 | 1.7 | [14] |
| Co/Pt (RT) | $5 \times 10^{11}$ | 0.035 | 0.025 | [16] |
| Ta/Co (RT) | $1.2 \times 10^{11}$ | 0.004 | > 1 | [14] |
| GaMnAs/BiSb (30K) | $1.5 \times 10^{10}$ | 1.1 | 0.2 | [26] |
| CBST/BST (4K) | $1.5 \times 10^{8}$ | 0.68 | 0.7 | [18] |